

\input harvmac
\noblackbox
\pageno=0\nopagenumbers\tolerance=10000\hfuzz=5pt
\line{\hfill CERN-TH.7265/94}
\vskip 36pt
\centerline{\bf DOUBLE ASYMPTOTIC SCALING AT HERA}
\vskip 36pt\centerline{Richard~D.~Ball\footnote{*}{On leave
from a Royal Society University Research Fellowship.}
 and Stefano~Forte\footnote{\dag}{On leave
from INFN, Sezione di Torino, Italy.}}
\vskip 12pt
\centerline{\it Theory Division, CERN,}
\centerline{\it CH-1211 Gen\`eve 23, Switzerland.}
\vskip 36pt
{\medskip\narrower
\ninepoint\baselineskip=9pt plus 2pt minus 1pt
\lineskiplimit=1pt \lineskip=2pt
\def\smallfrac#1#2{\hbox{${{#1}\over {#2}}$}}
\centerline{\bf Abstract}
\noindent
Perturbative QCD predicts that at sufficiently large $Q^2$ and
small $x$ nucleon structure functions should exhibit scaling in the
two variables $\sqrt{\ln\smallfrac{1}{x}\ln\ln Q^2}$
and $\sqrt{\ln\smallfrac{1}{x}\big/\ln\ln Q^2}$, provided only that the
small-$x$ behaviour of the input to the perturbative QCD evolution
is sufficiently soft. We derive these asymptotic
results by writing the gluonic Altarelli--Parisi equation at small
$x$ as a two--dimensional wave equation, which propagates the gluon
distribution from its boundaries into the asymptotic region.
We then show that the existing experimental data on $F_2^p(x,Q^2)$ from HERA
provide a remarkable confirmation of both of these scaling
predictions. The `hard' pomeron, which does not scale, is thereby
excluded by more than three standard deviations; only a very small
admixture of it is permitted by the data.
We propose that existing and future data from HERA should be binned
in the two scaling variables, in order to facilitate the search for
small scaling violations.
}
\vskip 16pt
\centerline{Submitted to: {\it Physics Letters B}}
\vskip 20pt
\line{CERN-TH.7265/94\hfill}
\line{May 1994\hfill}

\vfill\eject
\footline={\hss\tenrm\folio\hss}


\def\rhs{right hand side}
\def\lhs{left hand side}
\def\toinf#1{\mathrel{\mathop{\sim}\limits_{\scriptscriptstyle
{#1\rightarrow\infty }}}}
\def\frac#1#2{{{#1}\over {#2}}}
\def\half{\hbox{${1\over 2}$}}
\def\quarter{\hbox{${1\over 4}$}}
\def\smallfrac#1#2{\hbox{${{#1}\over {#2}}$}}

\def\GeV{{\rm GeV}}

\catcode`@=11 
\def\slash#1{\mathord{\mathpalette\c@ncel#1}}
 \def\c@ncel#1#2{\ooalign{$\hfil#1\mkern1mu/\hfil$\crcr$#1#2$}}
\def\lsim{\mathrel{\mathpalette\@versim<}}
\def\gsim{\mathrel{\mathpalette\@versim>}}
 \def\@versim#1#2{\lower0.2ex\vbox{\baselineskip\z@skip\lineskip\z@skip
       \lineskiplimit\z@\ialign{$\m@th#1\hfil##$\crcr#2\crcr\sim\crcr}}}
\catcode`@=12 

\def\SF{\sigma}\def\RDB{\rho}\def\mass{\gamma}
\def\PR{{\it Phys.~Rev.~}}
\def\PRL{{\it Phys.~Rev.~Lett.~}}
\def\NP{{\it Nucl.~Phys.~}}

\def\PL{{\it Phys.~Lett.~}}
\def\PRep{{\it Phys.~Rep.~}}
\def\AP{{\it Ann.~Phys.~}}

\def\SJNP{{\it Sov.~Jour.~Nucl.~Phys.~}}
\def\SPJETP{{\it Sov.~Phys.~J.E.T.P.~}}

\def\vol#1{{\bf #1}}\def\vyp#1#2#3{\vol{#1} (#2) #3}


The main evidence for perturbative QCD in deep inelastic scattering at
large $x$ comes from the observation of small violations of Bjorken
scaling \ref\Bj{J.D.~Bjorken, \PR\vyp{179}{1969}{1547}.}
in the structure function $F_2(x;t)$  in the Bjorken limit
of large $t\equiv\ln Q^2/\Lambda^2$ at fixed $x$. These scaling
violations may be accurately predicted using the operator product
expansion \ref\GPGW{H.~Georgi and H.D.~Politzer,
\PR\vyp{D9}{1974}{416}\semi D.~Gross and F.~Wilczek,
\PR\vyp{D9}{1974}{980}.} or
equivalently  the Altarelli--Parisi
equations \ref\AP{G.~Parisi, {\it Proc. 11th Rencontre de Moriond,}
ed. J.~Tran~Thanh~Van, ed. Fronti\`eres, 1976\semi
G.~Altarelli and G.~Parisi, \NP\vyp{B126}{1977}{298}\semi
G.~Altarelli, \PRep\vyp{81}{1981}{1}.}. In this letter we will
show that in the double limit of large $t$ and small $x$ perturbative
QCD also  makes definite scaling predictions, provided only that the
gluon distribution at scales of order 1~GeV is reasonably
soft at small $x$ (i.e. that the pomeron intercept is
close to one). The appropriate scaling variables are the
geometric mean $\SF\simeq\sqrt{\ln\smallfrac{1}{x}\ln t}$ and the ratio
$\RDB\simeq\sqrt{\ln\smallfrac{1}{x}\big/\ln t}$; the gluon distribution
then scales asymptotically in both $\SF$ and $\RDB$ in the
double limit of large $\SF$ at fixed $\RDB$ and large $\RDB$ at
fixed $\SF$. These simple
predictions, which are asymptotically parameter free up to an overall
normalization factor, may be directly tested by comparison with experiment.
Indeed, we will show that recent data from HERA
\nref\ZEUS{ZEUS~Collab.,
\PL\vyp{B316}{1993}{412}.}\nref\Hone{H1~Collab.,
\NP\vyp{B407}{1993}{515}.}\refs{\ZEUS,\Hone} are in remarkable
agreement with this asymptotic behaviour. Furthermore, since the
double asymptotic scaling would
necessarily be spoiled if the pomeron were hard,
we will be  able to use it as a means to
discriminate experimentally between soft and hard pomerons.

Before actually proceeding to the comparison of double
asymptotic scaling behaviour with the data, we first derive it
in a simple but, we believe, instructive way by
reducing the singlet Altarelli--Parisi equations at small-$x$ and large
$t$ to a wave--like equation for the gluon distribution, which
may be solved exactly. We then derive the asymptotic form of this
solution for large values of the scaling variables $\SF$ and $\RDB$,
and different classes
of boundary conditions, namely `soft' and `hard'. We explain in what
sense solutions with soft boundary conditions exhibit double
asymptotic scaling, establish the kinematic region in
which this scaling behaviour sets in, and show the way in which it is
corrupted if the boundary conditions are `hard'. We derive
a simple relation between $F_2$ and the gluon distribution in
the asymptotic region, and finally proceed to a direct comparison
with the experimental data.

We will be concerned specifically with the singlet contribution to the
proton structure function
\eqn\eFtwo{\eqalign{F_2^p(x;t)
&=\smallfrac{5}{18}xq(x;t)+F_2^{NS}(x;t)\cr
q(x;t)&\equiv\sum_{i=1}^{n_f}\big(q_i(x;t)+\bar{q}_i(x;t)\big),\cr}}
where $n_f$ is the number of active flavours.
At small enough $x$ the nonsinglet contribution $F_2^{NS}(x;t)$
is negligible and can be ignored.
At small $x$ and
large $t$ the singlet quark distribution $q(x;t)$ is essentially
driven by the generic instability of the gluon distribution $g(x;t)$;
determining the latter thus allows us to compute $F_2^p$ in this
kinematic region.

\nref\DLLA{A.~De~Rujula, S.L.~Glashow, H.D.~Politzer, S.B.~Treiman,
           F.~Wilczek and A.~Zee,\hfil\break \PR\vyp{D10}{1974}{1649}\semi
        see also Yu.L.~Dokshitzer, \SPJETP\vyp{46}{1977}{641}.}
\nref\sxmell{M.~Gl\"uck and E.~Reya, \PR\vyp{D14}{1976}{3034}\semi
F.~Martin, \PR\vyp{D19}{1979}{1382}.}
\nref\sxad{T.A.~DeGrand, \NP\vyp{B151}{1979}{485}.}
\nref\GLR{L.V.~Gribov, E.M.~Levin and M.G.~Ryskin
\PRep\vyp{100}{1983}{1}.}
To see how this works, consider the singlet Altarelli--Parisi equations \AP,
which describe perturbative evolution of
$g(x;t)$ and $q(x;t)$ in the kinematical region displayed in
\fig\fxt{Region of validity of the Altarelli-Parisi equation
          in the $(x,t)$ plane. The
          disallowed regions are $t<t_0$, where perturbation theory
          breaks down because $\alpha_s(t)$ becomes too large, and
          $x\lsim x_r\exp(-\alpha_s(t_0)^2/\alpha_s(t)^2)$, where
          recombination effects may become important \GLR. The
          small-$x$ evolution equations may be used above the dashed line.
          The dotted region with the ? is that in which the Lipatov
          equation may become applicable.}:
\eqn\eAP{\eqalign{\frac{\partial}{\partial t}g(x;t)
&=\frac{\alpha_s(t)}{2\pi} \int_x^1 \frac{dy}{y}
       [P_{gg}(\smallfrac{x}{y})g(y;t)
        + P_{gq}(\smallfrac{x}{y})q(y;t)],\cr
\frac{\partial}{\partial t}q(x;t)
&=\frac{\alpha_s(t)}{2\pi} \int_x^1 \frac{dy}{y}
       [2n_f P_{qg}(\smallfrac{x}{y})g(y;t)
        + P_{qq}(\smallfrac{x}{y})q(y;t)],\cr}}
where $\alpha_s(t)=\smallfrac{4\pi}{\beta_0 t}+O(\smallfrac{1}{t^2})$,
$\beta_0=11-\smallfrac{2}{3}n_f$. Solving these equations sums up all
leading logarithms of the form $\alpha_s^m\ln^m Q^2
\ln^n\smallfrac{1}{x}$ with $n\leq m$; terms with fewer powers of
$\ln Q^2$should be negligible if $t$ is large enough for perturbation
theory to be trusted ($t>t_0$ say, where $\alpha_s(t_0)\ll 1$).

Now evolution at small $x$ ($x<x_0$, say) will be dominated by the
lowest moments of the splitting functions; this may be demonstrated
explicitly by taking the Mellin transform of the evolution equations
\eAP, solving them, and then using a saddle-point evaluation of the
inverse Mellin transform of the solution\refs{\DLLA,\sxmell}.
In practice, this is equivalent to expanding the Mellin transform of
the one loop splitting functions about their largest singularity in
moment space (which is when the moment variable $n\to 0$) and
discarding terms which vanish as $n\to 0$\refs{\sxad,\GLR}.
Inverting the transform
again, this leads to the approximate  form of the splitting functions
\eqn\esplitsx{{\eqalign{P_{gg}(z)
&\approx\smallfrac{6}{z}-(\smallfrac{11}{2}+\smallfrac{n_f}{3})\delta(1-z),\cr
P_{qg}(z)&\approx\smallfrac{1}{3}\delta (1-z),\cr}}
\qquad{\eqalign{P_{gq}(z)&\approx\smallfrac{8}{3z}-2\delta (1-z),\cr
P_{qq}(z)&\approx 0.\cr}}}

When substituted into \eAP, the terms in \esplitsx\ which go as $z\inv$ will
sum all the double logarithms, while the terms proportional to
$\delta(1-z)$ will approximately sum the remaining single
logarithms\foot{This approximation could be refined by keeping higher
powers of $n$ in the Mellin transform of the splitting function. This
leads to further terms in \esplitsx\ involving derivatives of the
delta function.};
in terms of ladder diagrams, the former terms account for rungs which
are strongly ordered in both $x$ and $t$, the latter for rungs which
are only strongly ordered in $t$. If only the former terms were
retained one would get the so--called double--leading logarithm
approximation.

Defining for convenience two new variables
\eqn\exizeta{\xi\equiv\ln\left(\frac{x_0}{x}\right),
       \qquad\zeta\equiv\ln\left(\frac{t}{t_0}\right)
                       =\ln\left(\frac{\ln Q^2/\Lambda^2}
                                      {\ln Q_0^2/\Lambda^2}\right)}
(so that at $x=x_0$ $\xi=0$, and at $t=t_0$ $\zeta=0$), and writing
\eqn\eGQ{G(\xi,\zeta)\equiv xg(x;t),\qquad Q(\xi,\zeta)\equiv
xq(x;t)=\sum_i x\big(q_i(x;t)+\bar q_i(x;t)\big),}
then substituting \esplitsx\ into \eAP, and differentiating the first
equation with respect to $\xi$, the singlet evolution equations
assume the simple form
\eqnn\eAPsxg\eqnn\eAPsxq
$$\eqalignno{\frac{\partial^2}{\partial\xi\partial\zeta}
G(\xi,\zeta)+\delta'\frac{\partial}{\partial\xi}
G(\xi,\zeta)-\mass^2 G(\xi,\zeta)&=
\smallfrac{4\mass^2}{9}Q(\xi,\zeta)
-\smallfrac{\mass^2}{3}\frac{\partial}{\partial\xi} Q(\xi,\zeta),&\eAPsxg\cr
\frac{\partial}{\partial\zeta}Q(\xi,\zeta)&=\smallfrac{n_f\mass^2}{9}
G(\xi,\zeta),&\eAPsxq\cr}$$
where $\mass\equiv \sqrt{12/\beta_0}$ and
$\delta'\equiv(11+\smallfrac{2n_f}{3})\big/\beta_0$.
These equations should be valid throughout the
first quadrant of the $(\xi,\zeta)$ plane (
\fig\fxize{The ($\xi$,$\zeta$) plane, showing the backward light cone
          at the point ($\xi'$,$\zeta'$), curves of constant
          $\SF$ (the hyperbolae) and lines of constant $\RDB$.}),
outside the multiple rescattering region $\xi\gg e^{2\zeta}$ \GLR,
up to corrections of order
$x_0e^{-\xi}$ and $\alpha_s(t_0)e^{-\zeta}$.\foot{Corrections to
the expansions \esplitsx\ of the one loop
splitting functions will lead to contributions to the
parton distributions which are suppressed by powers of
$\rho\inv$ (where $\rho$ is as defined in (11) below). Furthermore
it may be shown by explicit computation that the two loop corrections to
\esplitsx\ have little effect on the final form of $F_2$ provided
$\sigma\gsim 1$ and $1\lsim\rho\lsim 3$. However for $n\lsim 0.35$,
which in practice means $\rho\gsim 3$, there are indications that
perturbation theory may begin to break down\ref\EKL{
R.K.~Ellis, Z.~Kunszt, and E.M.~Levin,
Fermilab-PUB--93/350-T, ETH-TH/93--41, to be published in \NP B.}.}

The inhomogeneous term on the \rhs\ of \eAPsxg\ is
negligible due to its small coefficient and may be dropped.
However, this  is not necessary: if the matrix of
splitting functions is diagonalized before the Mellin inversion,\foot{
Although in general the diagonalization introduces an extra
singularity (a branch cut) this is
not a problem here since we expand about $n=0$.} this
term is in effect absorbed into a small shift in $\delta'$, which
becomes $\delta\equiv (11+\smallfrac{2n_f}{27})\big/\beta_0$. The
gluon evolution equation \eAPsxg\ then becomes homogeneous,
\eqn\ewave{\Big[\frac{\partial^2}{\partial\xi\partial\zeta}
+\delta\frac{\partial}{\partial\xi}-\mass^2 \Big] G(\xi,\zeta)=0,}
while the quark evolution equation \eAPsxq\ remains unchanged (up
to subleading corrections).


We next make the essentially
trivial observation that \ewave\ is in fact a two dimensional
wave equation, with  `time' $\half(\xi+\zeta)$ and
`space' $\half(\xi-\zeta)$). This allows us to immediately infer
many useful properties of its solutions, and thus deduce directly many
well--known
qualitative features of parton evolution. In particular:
\item{(i)} The equation is essentially symmetrical in $\xi$ and $\zeta$, so
the gluon distribution $xg(x;t)$ evolves (`propagates') equally
in both $x$ and $t$ (up to the (small) asymmetry induced by the
`damping' term proportional to $\delta$). Any further
asymmetry in $\xi$ and $\zeta$ must thus come
from the boundary conditions.
\item{(ii)} The propagation is `timelike', into the forward `light-cone'
at the origin $(\xi,\zeta)=(0,0)$, along the `characteristics'
$\xi={\rm constant}$ and $\zeta={\rm constant}$ (see \fxize).
\item{(iii)} At a given point $(\xi,\zeta)$, $G(\xi,\zeta)$ depends
only on the boundary conditions contained within
the backward light cone formed by the two characteristics through
$(\xi,\zeta)$ \fxize. This means that to calculate $xg(x;t)$ it is
unnecessary to know what happens at smaller values of $x$ or larger
values of $t$.
\item{(iv)} Because the equation is linear, contributions to
$G(\xi,\zeta)$ from different parts of the boundaries are simply
added together.
\item{(v)} Since the `mass' term is negative
($\mass^2>0$), the propagation is `tachyonic'; this means that $G(\xi,\zeta)$
is unstable, growing exponentially rather than oscillating.
\item{(vi)} Since $\delta>0$ the damping term ensures that, at fixed
$\xi$ $G(\xi,\zeta)$ eventually falls with increasing $\zeta$.

Setting $G(\xi,\zeta)$ along two characteristics (for
example $\xi=0$ and $\zeta=0$) gives a well--posed problem (the
`characteristic Goursat problem' \ref\Goursat
{E.~Goursat, {\it Cours d'Analyse Math\'ematique,
                            Vol.III}, Gauthier--Villars, 1902.})
provided only that the boundary
conditions are compatible at the point where the boundaries meet.
Since the Green's function for the wave operator \ewave\ may be
expressed in terms of the Bessel function
\eqn\ebess{I_0(z)\equiv\sum_0^\infty\frac{\big(\quarter z^2\big)^n}{(n!)^2}
                 \toinf{z}\frac{1}{\sqrt{2\pi z}}e^z
                             \big(1+O(\smallfrac{1}{z})\big),}
the general solution to this problem is simply
\eqn\eGoursat{\eqalign{G(\xi,\zeta)=
            I_0\big(2\mass\sqrt{\xi\zeta}\big)&e^{-\delta\zeta}G(0,0)
            +\int_0^\xi d\xi' I_0\big(2\mass\sqrt{(\xi-\xi')\zeta}\big)
            e^{-\delta\zeta}\frac{\partial}{\partial\xi'}G(\xi',0)\cr
           &+\int_0^\zeta d\zeta' I_0\big(2\mass\sqrt{\xi(\zeta-\zeta')}\big)
                                   e^{\delta(\zeta'-\zeta)}
 \Big(\frac{\partial}{\partial\zeta'}G(0,\zeta')+\delta G(0,\zeta')\Big).\cr}}
%


The asymptotic behaviour of $G$ away from the boundaries will depend
in general on whether the boundary conditions are compact or
noncompact, i.e., on whether the starting distributions are soft or
hard at large $\xi$ on the $\zeta=0$ boundary and large $\zeta$ on the
$\xi=0$ boundary. Due to the linearity property, we can consider each boundary
separately, and then add the various contributions.
We consider first the simplest case in which both boundary conditions
are soft, so $g(x;t_0)\sim x\inv$ as $x\to 0$ (i.e. $G(\xi,0)\sim
{\rm const.}$ as $\xi\to\infty$), the traditional
behaviour inferred from the intercept of the `soft' pomeron
\ref\softpom{H.D.I.~Abarbanel, M.L.~Goldberger and S.B.~Treiman,
\PRL\vyp{22}{1969}{500}\semi P.V.~Landshoff, J.C.~Polkinghorne and
R.D.~Short, \NP\vyp{B28}{1970}{210}.}, while $g(x_0;t)\sim t^{-\delta}$ as
$t\to\infty$ (i.e. $G(0,\zeta)\sim e^{-\delta\zeta}$ as $\zeta\to\infty$).
Since in this case `waves' are only
generated close to the origin, far from the origin we may
determine the behaviour of the solution \eGoursat\ by means of a
multipole expansion, i.e., expanding the Bessel
functions under the integrals in eq.\eGoursat\ in powers of
$\smallfrac{\xi'}{\xi}$ and $\smallfrac{\zeta'}{\zeta}$ respectively.
Defining the new scaling variables
\eqn\esr{\SF\equiv\sqrt{\xi\zeta}
=\sqrt{\ln\smallfrac{x_0}{x}\ln\smallfrac{t}{t_0}},
\qquad\RDB\equiv\sqrt{\frac{\xi}{\zeta}}=
\sqrt{\frac{\ln\smallfrac{x_0}{x}}{\ln\smallfrac{t}{t_0}}},}
and considering the limit $\SF\to\infty$
at fixed (but reasonably large) $\RDB$ (see \fxize),\foot{More precisely,
we consider the limit $\SF\to\infty$ along any
curve such that also $\RDB\to\infty$, such as
for example the curve $\xi\propto \zeta^{1+\epsilon}$ with $\epsilon>0$.}
we then find asymptotically the generic behaviour
\eqnn\euniv
$$\eqalignno{G(\SF,\RDB)
      &\toinf{\SF} I_0(2\mass\SF)
            e^{-\delta({\SF/\RDB})}
          \bigg\{G(0,0)&\cr
       &\qquad +\int_0^\xi \!d\xi'\frac{\partial G}{\partial\xi'}
            e^{-(\mass/\RDB)\xi'}
              + \int_0^\zeta \!
            d\zeta'\bigg(\frac{\partial G}{\partial\zeta'}+\delta G\bigg)
            e^{(\delta-\mass\RDB)\zeta'}
             +     O(\smallfrac{1}{\SF})\bigg\} &\cr
      &\toinf{\SF}
          {\cal N} f_g(y)\frac{1}{\sqrt{4\pi\mass\SF}}
       \exp\Big\{ {2\mass\SF -\delta\big(\smallfrac{\SF}{\RDB}\big)}\Big\}
      \big(1+O(\smallfrac{1}{\SF})\big).&\euniv\cr}$$
where $y\equiv\mass/\RDB$, and in the second line we used the
asymptotic behaviour \ebess\ of
the Bessel function. The normalization factor ${\cal N}$ is just that
of the boundary condition\foot{The boundary condition $g(x;t_0)$ will be
normalized so that the second moment (the gluon momentum
fraction) is unity; the boundary condition $g(x_0,t)$ is then
normalized by continuity at $(x_0,t_0)$. When comparing to the data,
the overall normalization will thus give the gluon momentum
fraction.}, while $f_g(0)=1$, but $f_g(y)$ will
depend on the particular details of the boundary condition,
while being generally rather smooth.\foot{For example if
$xg(x;t_0)= {\cal N} (1-x)^\beta$, $x_0g(x_0;t)= {\cal N}
(1-x_0)^\beta\big(\smallfrac{t}{t_0}\big)^{-\delta}$,
then $f_g(y)=(1-x_0)^\beta
+\beta x_0\sum_{n=0}^{\beta-1} \smallfrac{(\beta -1)!}{n!(\beta-1-n)!}
\smallfrac{(-x_0)^n}{1+n+y}\sim 1-\beta x_0{y\over1+y}+O(x_0^2)$ as
$x_0\to 0$.}
The asymptotic behaviour \euniv\ is thus, as expected, independent of
the detailed form of boundary condition, up to corrections of order
$1/\RDB$, which are anyway  of the same order as corrections
due to the terms lost by the truncation \esplitsx\ of the expansion of
the splitting functions.

If instead the boundary conditions are noncompact, it is not
difficult to see from \eGoursat\ that the form of $G(\xi,\zeta)$ on
the boundary will tend to persist into the interior of the
$(\xi,\zeta)$ plane; the asymptotic behaviour close to the boundary
will then not be universal, but depend on the particular boundary
conditions adopted.
Consider for example $xg(x;t_0)\sim {\cal N}^\prime x^{-\lambda}$,
(phenomenologically $\lambda\simeq 0.08$ \ref\pomint{
P.D.B.~Collins and F.~Gault, \PL\vyp{112B}{1982}{255}\semi
A.~Donnachie and P.V.~Landshoff, \NP\vyp{B244}{1984}{322};
 \NP\vyp{B267}{1986}{690}.}).
Then if $\RDB>\frac{\mass}{\lambda}$, the first boundary integral
in eq.\eGoursat\ is dominated by a nontrivial saddle-point,
and gives the asymptotic behaviour
\eqn\ehpa{G(\SF,\RDB)\toinf{\SF} {\cal N}^\prime
\exp{\big\{\lambda\SF\RDB +
(\smallfrac{\mass^2}{\lambda}-\delta)(\smallfrac{\SF}{\RDB})\big\}}
      \big(1+O(\smallfrac{1}{\SF})\big),}
to be added to \euniv. Very close to the boundary ($\RDB\gsim 15$) the
power--like growth at small $x$ is thus preserved by the evolution to
$t>t_0$. However further away from the boundary, in the region
$\RDB\lsim\frac{\mass}{\lambda}$ the more gentle growth \euniv\
is unchanged, modulo a slight increase in ${\cal N}$.
Similarly, if the boundary condition on the other boundary is hard
(say $G(0,\zeta)\sim e^{-\delta'\zeta}$ as $\zeta\to\infty$, with
$0\lsim\delta'<\delta$), this too will propagate into the region
$\RDB\lsim\frac{(\delta-\delta')}{\mass}$, in effect reducing the
value of $\delta$ close to the boundary. Again away from the boundary
the universal behaviour \euniv\ is unaffected.

Thus with only the rather conservative assumption of reasonably soft
boundary conditions (i.e. for sufficiently small $\lambda$ and
$\delta'$) perturbative QCD predicts a universal
growth \euniv\ in the gluon structure function at large $t$ and
small $x$, faster than any power of $\ln\smallfrac{1}{x}$ but slower
than any inverse power of $x$. This means that away from the boundaries
perturbative QCD allows us to predict in a
parameter--free way (i.e. independently of the precise form of the
boundary conditions) the asymptotic behaviour of the gluon
distribution, and hence (as we will see below) of the $F_2$ structure
function, throughout a very large region of the $(x,t)$ plane \fxt.
Although it has been known now for twenty years \DLLA, hitherto this
remarkable consequence of the perturbative QCD evolution equations seems
not to have been taken very seriously. Here we hope to remedy this.

We begin by noting that the universal asymptotic behaviour \euniv\
actually implies scaling in both $\SF$ and $\RDB$: $\ln G(\SF,\RDB)$  at
fixed $\RDB$ is asymptotically a linear function of $\SF$, with slope
independent of $\RDB$ (up to terms which vanish as $1/\RDB$), while at
fixed $\SF$ it is asymptotically a flat function of $\RDB$. In other
words $\SF\inv\ln G(\SF,\RDB)$ is asymptotically independent of both
$\SF$ and $\RDB$, up to corrections which vanish as $1/\SF$ and
$1/\RDB$. This `double asymptotic scaling' is a
direct result of two generic properties of wave propagation from
compact boundaries: at large distances from the boundaries, the details of
the boundary conditions are washed out ($\SF$ scaling)\foot{The
position of the boundaries is also unimportant
asymptotically, so even if $t_0$ is taken to be very small
the same generic behaviour will be found at small $x$
and large $t$; as we will see, this is sufficient to explain the
success of the predictions of ref.\ref\GRV{
M.~Gl\"uck, E.~Reya and A.~Vogt, \PL\vyp{B306}{1993}{391} and
ref. therein.}.}, and, apart from
the damping term (which is asymptotically subleading) the propagation
is isotropic ($\RDB$ scaling).

This double scaling property is peculiar to the universal behaviour
\euniv\ derived from reasonably soft boundary conditions; instead, if one of
the boundary conditions is very hard, its influence will spread into a
large part of the $(\xi,\zeta)$ plane, spoiling both $\RDB$ and $\SF$
scaling in this region in a way that will depend explicitly on the
particular form of the hard boundary condition. For example consider
the Lipatov `hard' pomeron\ref\hardpom{
        L.N.~Lipatov, \SJNP\vyp{23}{1976}{338}\semi
          V.S.~Fadin, E.A.~Kuraev and L.N.~Lipatov,
       \PL\vyp{60B}{1975}{50};
       {\it Sov. Phys. JETP~}\vyp{44}{1976}{443};\vyp{45}{1977}{199}\semi
          Y.Y.~Balitski and L.N.Lipatov, \SJNP\vyp{28}{1978}{822}.},
which grows as a power of $\smallfrac{1}{x}$ as $x\to 0$ at fixed $t$:
$xg(x;t_0)\sim {\cal N}^\prime x^{-\lambda}$, with
$\lambda\lsim\smallfrac{12\ln 2}{\pi}\alpha_s(t_0)\approx\half$.
This behaviour is supposed to be valid at some intermediate $t$: it is
obtained by summing leading logarithms of the form $\alpha_s^n
\ln^n\smallfrac{1}{x}$, terms containing powers of $\ln Q^2$ being
ignored; $t$ must be small, but not so small that $\alpha_s(t)$ is too
large. It will thus serve as a boundary condition for Altarelli--Parisi
evolution into the region $t>t_0$. It will then lead to power--like
growth of the form \ehpa\ throughout the region $\RDB\gsim
\frac{\mass}{\lambda}$, which must be {\it added} to the weaker growth
\euniv\ over which it will dominate at large $\RDB$. Although the
structure function still grows at small $x$, it now increases so
strongly that double scaling is spoiled.

In what remains of this letter we will compare the double
asymptotic scaling prediction \euniv\ with all the available small-$x$
data from the two HERA experiments\refs{\ZEUS,\Hone}, to see whether
it is supported empirically, or whether there is instead evidence for
other less conventional physics (and in particular the `hard' pomeron)
at small-$x$. Firstly we must establish that there are data in the
asymptotic region where double scaling might apply.
We thus begin by examining the approach to asymptopia
as the general solution \eGoursat\ evolves away from the
boundaries. We choose $x_0=0.1$, which
should be sufficiently small for \esplitsx\ to be a good
approximation. The starting scale is chosen as  $Q_0=1\GeV$:
not so small that perturbative evolution has broken down, but
sufficiently small that the behaviour at small $x$ is given by a
nonperturbative boundary condition (or, in the case of the hard
pomeron, sufficiently small that logarithms of $Q_0^2$ may be
reasonably ignored).

\nref\NMC{NMC~Collab., \PL\vyp{B295}{1992}{159}.}
In
\fig\fcontours{Propagation of $G$ into the $\xi$-$\zeta$ plane;
              a) soft pomeron, b) hard pomeron. A scatter plot of the
              NMC and HERA data is superimposed for reference:
              the two types of crosses are NMC data \NMC\ (taken with
              90~GeV and 280~GeV beams respectively), the diamonds
              are  ZEUS data \ZEUS, and the squares  H1 data \Hone.}
we display a contour plot of $G(\xi,\zeta)$ computed from the solution
\eGoursat\ with, for definitness, the soft boundary condition
$G(\xi,0)={\cal N}(1-x_0e^{-\xi})^9$ compared with the solution
computed from the hard boundary condition $G(\xi,0)={\cal N}^\prime
x_0^{-1/2}e^{\xi/2}(1-x_0e^{-\xi})^9$. On the $\xi=0$
boundary we take in both cases $G(0,\zeta)=G(0,0)$
as (approximately) required empirically when $x_0=0.1$.
The plots clearly show that the asymptotic behaviour is attained very
rapidly in the soft case, around $\SF\sim 1$; the damping term and the
propagation from the lower boundary distort the curves somewhat at
small values of $\RDB$, as expected.
In the hard case the asymptotic behaviour \ehpa\ clearly holds only for
large $\RDB\gsim 3$, while in most of the rest of the region
the full solution is a compromise between the two sorts of behaviour.
In practice, this means that in much of the HERA kinematic range the
evolved hard solution is actually rather similar to the
unevolved boundary condition imposed at $\zeta=0$.

It is now clear that, because almost all of
the HERA data (and even some of the older NMC data) lie in the region
$\SF\in[1,2]$, $\RDB\in[1,3]$, well
inside the asymptotic kinematic regime of the soft solution,
the universal behaviour \euniv\ should be easily tested, while the
hard behaviour should be easily
distinguished from it. This is demonstrated explicitly in
\fig\fconv{Scaling plots of the function $G(\SF,\RDB)$ displayed
           in \fcontours. The solid lines correspond to \fcontours a)
           (the soft pomeron) and the dashed lines to \fcontours b)
           (the hard pomeron); the dotted lines correspond to an
           unevolved hard initial condition.\hfill\break
           i) $\ln G(\SF,\RDB)$ is plotted versus $\SF$; the lower line
           of each pair has $\RDB=1.4$ and the upper one
           $\RDB=3.2$.\hfill\break
           ii)The rescaled function  $R_G(\SF,\RDB)G(\SF,\RDB)$ is
           plotted versus $\RDB$; the lower line
           of each pair has $\SF=1.2$ and the upper one $\SF=2.1$.}
where we compare the scaling expected from the soft pomeron
boundary condition with the hard pomeron prediction (with $\lambda=\half$).
Not only does the hard pomeron generally lead to more gluons in any given
range of $x$, but the number rises rather more steeply at smaller
values of $x$. Thus at fixed $\RDB$ (\fconv i), even though $\ln G(\SF,\RDB)$
is in both cases an (approximately) linear function of $\SF$ in the
hard case its slope gets steeper for larger values of $\RDB$, while in
the scaling case it is universal. Correspondingly at fixed $\SF$, in the
soft case $G(\SF,\RDB)$ becomes flat as $\RDB$ grows, while in the
hard case it rises rather rapidly.
This is best exhibited
by first performing a simple rescaling of $G(\SF,\RDB)$, by a factor
\eqn\rescg
{R_G(\SF,\RDB)=\exp(-2\mass\SF+\delta(\SF/\RDB)+\half\ln\SF),}
so that the curves with different values of $\SF$ lie on top of one
another (\fconv ii). Besides showing the clear distinction to be made
between the predictions derived from soft and hard boundary
conditions, these plots also
confirm that for compact boundary conditions, the asymptotic double
scaling regime is indeed reached rather rapidly, at moderately small
values of $\SF$ and $\RDB$.


Before coming finally to a direct comparison of this predicted
behaviour to the data \refs{\ZEUS,\Hone} we must still relate
the calculated gluon distribution to the proton structure function
$F_2^p$ measured in the experiments. This is actually rather simple
at small $x$ and large $t$, since all that is necessary is to
integrate \eAPsxq:
\eqn\eFfromG{F_2^p=\smallfrac{5}{18}Q(\xi,\zeta)=\smallfrac{5}{18}Q(\xi,0)
+\smallfrac{5n_f}{162}\mass^2\int_0^\zeta \!d\zeta' G(\xi,\zeta').}
With compact boundary conditions, the asymptotic form of $F_2^p$ may
then be deduced directly from that of $G$, by substituting \euniv\ into
\eFfromG, to give (whenever $\RDB >\delta/\mass$)
\eqn\eFasymp{F_2^p\toinf{\SF}
    \smallfrac{5n_f}{162}y
    f_q(y)G(\SF,\RDB)\big(1+O(\smallfrac{1}{\SF})\big),}
where again $y={\mass\over\RDB}$, and
$f_q(y)\to 1$ as $y\to 0$.\foot{In fact the function $f_q$
may be determined by
integrating the second Altarelli--Parisi equation \eAP\ with
the second term set to zero (since its contribution is subasymptotic)
and $P_{qg}(z)=\half\big[z^2 + (1-z)^2\big]$; this gives
$f_q(y)=(1-\smallfrac{\delta}{\mass^2}y)\inv
\smallfrac{3(4+3y+y^2)}{2(1+y)(2+y)(3+y)}$. Since we already have a
similar uncertainty from $f_g$ in \euniv, due both to our
small $x$ approximation \ewave\ to the gluon evolution equation and
ignorance of the boundary conditions, this result will not be used
here, though it may be useful for extracting the gluon distribution
directly from the data.} A similar relation holds for the asymptotic
behaviour \ehpa, but with  $y=\lambda$  when
$\RDB>{\mass\over\lambda}$. So we
can actually use \eFasymp\ as a good approximation throughout the
asymptotic region for any type of boundary condition if we
set $y=\sqrt{(\mass/\RDB)^2+\lambda^2}$.
Apart from the subasymptotic factor of ${\mass/\RDB}$ the asymptotic growth
of the gluon distribution drives a similar growth of the singlet quark
distribution, and thus of the structure function $F_2^p$.


Using \eFasymp\ to determine a slightly different rescaling factor
\eqn\rescf
{R_F(\SF,\RDB)=\Big[\smallfrac{5n_f}{162}\smallfrac{\mass}{\RDB}\Big]\inv
                  R_G(\SF,\RDB).}
we may thus compare the $\SF$ and $\RDB$ scaling
plots \fconv\ to all the available data from ZEUS \ZEUS\
and H1 \Hone. The result is shown in
\fig\fscaling{Comparison between the experimental data and the
              predictions for $F_2(\SF,\RDB)$
              computed from $G(\SF,\RDB)$ displayed in \fconv,
              with normalization now adjusted according to the best--fit
              values given in the table. The diamonds indicate data
              from  ZEUS \ZEUS\ and the squares  data from
              H1 \Hone.\hfill\break
              i) The curves show $F_2$ computed from $G$ as
              displayed in \fconv i)  using eqn. \eFasymp; the dashed line is a
              linear fit to the HERA data; a) is the soft pomeron prediction
              (solid curves of \fconv i)) and b) is the hard pomeron
              (dashed curves of \fconv i)).
              The subasymptotic data from the NMC are also shown (crosses).
              The filled points correspond to data with
              $\RDB < 1.4$.\hfill\break
              ii) The curves are now calculated from $G$ as displayed
              in \fconv ii), compared to the HERA data rescaled by
              $R_F$ eq.\rescf. Again a) is the soft
              pomeron prediction and b) is the hard pomeron.}.
It should be immediately apparent that the agreement with the
asymptotic behaviour \euniv\ derived from the soft pomeron boundary
condition at $t_0$ is
remarkably good, especially when it is remembered that the asymptotic
prediction is parameter free, apart from the precise value
of the normalization. The hard pomeron boundary condition on the
other hand gives results which are a long way from the data, even
when its normalization is reduced.
Furthermore, the $\SF$ plot provides a nontrivial test of perturbative
QCD, since the fitted slope agrees very well with the computed
one. Also included in \fscaling i are the data from the NMC
collaboration \NMC; remarkably those which fall
within the asymptotic region agree very well with the line fitted to the
HERA data alone, despite their much smaller errors.

\topinsert\hfil
\vbox{\tabskip=0pt \offinterlineskip
      \def\tablerule{\noalign{\hrule}}
      \halign to 350pt{\strut#&\vrule#\tabskip=1em plus2em
                   &\hfil#\hfil&\vrule#
                   &#\hfil&\vrule#
                   &#\hfil&\vrule#
                   &\hfil#&\vrule#\tabskip=0pt\cr\tablerule
      &&\omit&&\omit\hidewidth $N_s$\hidewidth
             &&\omit\hidewidth $N_h$\hidewidth
             &&\omit\hidewidth $\chi^2$\hidewidth&\cr\tablerule
   &&   a)  && $0.33\pm 0.01$ && 0             &&  18~~(52) &\cr\tablerule
   &&   b)  && 0              && $0.14\pm0.01$ && 114~(282) &\cr
   && a)+b) && $0.31\pm 0.03$ && $0.01\pm0.01$ &&  17~~(50) &\cr\tablerule
   &&   c)  && 0              && $0.36\pm0.01$ && 121~(285) &\cr
   && a)+c) && $0.30\pm 0.05$ && $0.03\pm0.01$ &&  17~~(49) &\cr\tablerule
   &&   d)  && 0              && $0.17\pm0.01$ &&  67~(168) &\cr
   && a)+d) && $0.30\pm 0.05$ && $0.02\pm0.01$ &&  17~~(50) &\cr\tablerule}}
\hfil\bigskip
\centerline{\vbox{\hsize= 380pt \raggedright\noindent\footnotefont
Table: The fitted normalizations $N_s$ and $N_h$ and
the associated $\chi^2$s (those in parenthesis using only the
statistical errors in the experimental data).
The different cases considered are a) soft pomeron b)
hard pomeron c) hard pomeron (unevolved) d) `intermediate' pomeron,
and then the linear combinations a) + b), a) + c) and a) + d).
}}
\bigskip
\endinsert

We may assess quantitatively the quality of the fit of the predictions to
the HERA data by computing the $\chi^2$ of the hard and soft structure
functions $(F_2^p)^{\rm soft}$ and $(F_2^p)^{\rm hard}$
(obtained using \eFasymp\ from the gluon distributions displayed
in  \fcontours a) and b) respectively),
with the overall normalization of each left as a free parameter. The
results are displayed in the table. The $\chi^2$ of the
soft pomeron is rather good and fully compatible with the experimental
errors in the data: there are 38 data points, with presumably some
correlation in their systematic errors, and our $\chi^2$ is indeed
rather low if we combine the statistical and systematic errors for
each point in quadrature, but somewhat high if we use only the statistical
errors. The fit of the hard pomeron to the data is in contrast very
poor, excluding it at the level of at least five standard deviations.
Notice that the best-fit value of the normalization is in fact the
fraction of the nucleon's momentum carried by the gluons at the
starting scale $\zeta=0$; the value found in the case of soft boundary
conditions is thus in good qualitative agreement with phenomenological
expectations, whereas that for the hard boundary condition is
typically too small by a factor of two. As a cross--check, we also fitted
the linear combination $N_s(F_2^p)^{\rm soft}+N_h(F_2^p)^{\rm hard}$,
so that both the absolute and relative normalizations of the hard and
soft components are left as free parameters. As may be seen from the
table, only a very small admixture (less that about $5\%$) of the
hard component is permitted.

We have also determined the corresponding results for an unevolved hard
pomeron (again with intercept $\lambda=\half$). This would
presumably be appropriate if the Lipatov equation were somehow to remain valid
even when $t\gg t_0$, i.e. in the HERA region.\foot{Amusingly this
unevolved hard pomeron may be imitated by adding a negative damping term
$-\lambda\partial G/\partial\zeta$ to the \lhs\ of the gluon evolution
equation \ewave, thereby including $\alpha_s^n\ln^n\smallfrac{1}{x}$
terms in the evolution, and then evolving the soft pomeron from
$t=t_0$.}. It may be seen from the table that in fact this makes very
little difference, the improvement in fit as compared to the evolved
hard pomeron being at best marginal. Since these are two extreme
scenarios, they
bracket the intermediate possibility that the value $t_0$ at which the
Lipatov boundary condition should be imposed is somewhere in the middle
of the HERA region. In the same spirit we also include an
`intermediate' pomeron with intercept $\lambda=0.35$, since this is supposed
to be the minimum value expected from the Lipatov equation \ref\CK{
J.C.~Collins and J.~Kwiecinski, \NP\vyp{B316}{1989}{307}.}. Although
this fits a little better than the hard pomeron, it is still clearly excluded.

Finally, we have checked that the $\chi^2$ of the soft pomeron
solution is not changed significantly if the value of $t_0$ is
reduced, or if if the value of
$\Lambda$ is varied within present experimental uncertainty.  Of
course, if $t_0$ is raised too much, so that the data are not yet
in the asymptotic region of the soft solution, then $\chi^2$
starts growing and becomes almost as bad as that of the unevolved hard
solution: while the data clearly exhibit asymptotic double scaling, this
necessarily implies a growth at small $x$ for fixed $t$ which could
easily be misinterpreted as evidence for the hard pomeron.

We conclude that there is simply no room for more than a very small
admixture of the hard pomeron when $x\gsim 10^{-4}$.
On the contrary, the data agree surprisingly well with the universal double
scaling behaviour which only applies in the case of reasonably soft
boundary conditions. These results seem thus to agree with the
recent observation \ref\LC{J.C.~Collins and P.V.~Landshoff,
\PL\vyp{B276}{1992}{196}.} that if the integrated $k_T$ in the
Lipatov equation is cut--off in the ultraviolet, as
required kinematically, $\lambda$ is very significantly reduced for
reasonable values of $t$ , and it becomes very difficult to
distinguish the hard perturbative pomeron from the soft nonperturbative
one.

Since all existing published data is binned in $x$ and $Q^2$, it was
necessary when drawing our scaling plots \fscaling\ to use all the
data in each of the plots, rescaling with the function $R_F$ to ensure
that if the data exhibited scaling, they would lie on a single
straight line in each of the plots. A rather better test of asymptotic
double scaling could be made if the raw data were binned in $\RDB$ and
$\SF$: plots of $\ln F_2^p(\SF,\RDB)$ at fixed $\RDB$ and $\SF$
could then be drawn
to display both $\RDB$-scaling and $\SF$-scaling. Whereas the approach
to a flat ($\RDB$-independent) line of the plots at fixed $\SF$ would
indicate that asymptotic scaling has been attained, the value of the
asymptotic slope of the plots at fixed $\RDB$ would be given by the
universal value $2\mass=12\big/\sqrt{33-6n_f/N_c}$, a parameter-free
prediction of perturbative QCD which depends only on the number of
colours and flavours.

On the other hand, as much more data
becomes available, it should become possible to sharpen the tests of
double scaling to the point where the subasymptotic scaling violations
at small $\SF$ and $\RDB$ can be observed and measured; this would be
interesting as some of the leading violations (those which are
included in \rescg) are universal, while the subleading violations
could be calculated quite precisely using the known parton distributions
at large $x$ and the full evolution equations. It should then be
possible to extract (in a rather clean way) $\alpha_s$ from the HERA data.
More excitingly, one could also search systematically for scaling violations
at large $\SF$ and $\RDB$ due to new effects such as the hard pomeron
or parton recombination.\foot{Indeed the $\RDB$-scaling plot
\fscaling b) does seem to indicate a small (but not as yet
statistically significant) rise for $\RDB\gsim 3$.} The accurate
determination of structure
functions at small $x$ will then not only provide a precise and highly
nontrivial test of perturbative QCD, but could also be used to search
for new physics at small $x$.

\medskip
{\bf Acknowledgements:} We would like to thank M.~Arneodo and V.~del~Duca for
discussions, R.G.~Roberts and R.K.~Ellis for correspondence, and
I.J.R.~Aitchison, P.H.~Damgaard and D.Z.~Freedman for encouragement.


\listrefs
\listfigs
\end